# Resistive Switching and Voltage Induced Modulation of Tunneling Magnetoresistance in Nanosized Perpendicular Organic Spin Valves

Robert Göckeritz,[1] Nico Homonnay,[1] Alexander Müller,[1] Bodo Fuhrmann,[2] and Georg Schmidt[1,2,a]

[1]Institut für Physik, Martin-Luther-Universität Halle-Wittenberg, 06099 Halle (Saale), Germany
[2]Interdisziplinäres Zentrum für Materialwissenschaften, Martin-Luther-Universität Halle-Wittenberg, 06099 Halle (Saale), Germany
[a]Electronic mail: georg.schmidt@physik.uni-halle.de.

Nanoscale multifunctional perpendicular organic spin valves have been fabricated. The devices based on an $La_{0.7}Sr_{0.3}MnO_3$/Alq3/Co trilayer show resistive switching of up to 4-5 orders of magnitude and magnetoresistance as high as -70% the latter even changing sign when voltage pulses are applied. This combination of phenomena is typically observed in multiferroic tunnel junctions where it is attributed to magnetoelectric coupling between a ferromagnet and a ferroelectric material. Modeling indicates that here the switching originates from a modification of the $La_{0.7}Sr_{0.3}MnO_3$ surface. This modification influences the tunneling of charge carriers and thus both the electrical resistance and the tunneling magnetoresistance which occurs at pinholes in the organic layer.

In the past years a number of multiferroic tunnel junctions have been demonstrated in which tunneling magnetoresistance (TMR) and total device resistance can be modulated by a voltage pulse.[1,2] The effects are typically explained by tunneling electroresistance (TER) due to a ferroelectric barrier which changes the total resistance and magnetoelectric coupling at the interface between ferroelectric barrier and ferromagnetic contact which changes the TMR in magnitude and sometimes in sign. We observe the same functionality in organic spin valves (OSVs, Fig. 1), which after applying a voltage pulse may change the device resistance by three orders of magnitude or more and modulate their magnetoresistance (MR) from +26% to -38% which is a much larger effect than observed in Refs. 1 or 2. Nevertheless, the absence of a ferroelectric layer in our devices excludes both TER and magnetoelectric coupling as possible explanation. It should, however, be noted that our devices and those from Refs. 1 and 2 have a $La_{0.7}Sr_{0.3}MnO_3$ (LSMO) bottom electrode as a common property.

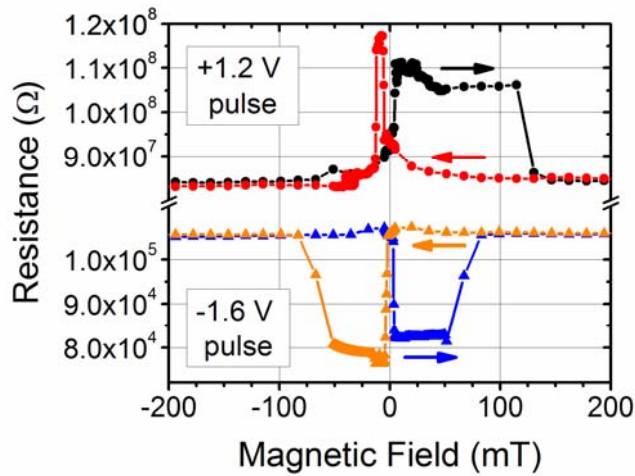

FIG. 1: MR traces of a nanosized OSV (LSMO/Alq3/MgO/Co/Ru with 20/12/3/30/10 nm in thickness) for two different resistance states after different voltage pulses taken at 4.3 K. The resistance changes by approx. three decades and the relative MR exhibit a sign reversal from +26% to -38%.

Already in 2011 the simultaneous observation of magnetoresistance and resistive switching (RS) has been reported for organic spin valves by Prezioso et al.[3,4] In this case the devices were LSMO/Alq3/Co-based spin valves showing a relative MR of -20% in the initial state. By applying voltage pulses the overall device resistance was increased while the relative MR decreased without changing shape or sign. The device resistance could be increased by two decades while the MR was completely suppressed. A possible explanation suggested by Prezioso *et al.* was the blocking of filaments or charge trapping in combination with giant magnetoresistance (GMR) and spin injection as a prevalent transport mechanism, however, no clear identification of the underlying physics was possible. Recent results from our own group in structures with only one ferromagnetic electrode (LSMO) also demonstrated RS. However, in this case the RS could clearly be linked to the modification of the LSMO surface which creates and modifies a tunnel barrier between the LSMO and the organic semiconductor.[5] The modification was shown to originate from the creation of mobile oxygen vacancies in the LSMO. In these experiments the MR could clearly be identified as tunneling anisotropic magnetoresistance (TAMR). Any increase in device resistance also resulted in increasing MR with a maximum value of 20%.

Similar to the OSV from Prezioso *et al.* the devices presented here have a second magnetic electrode. In recently published experiments we have already identified the origin of the magnetoresistance in our devices as tunneling via pinholes through the Alq3 layer resulting in TMR.[6,7] Also with respect to the interplay of RS and MR they differ considerably from the OSVs by Prezioso *et al.* as we do not only observe a change in magnitude but also a sign change of the MR.

All samples are fabricated using our recently reported technique[7] which allows to define perpendicular OSVs with nanosized active device area. The samples consist of 7 to 14 devices, respectively, with a layer stack of LSMO/Alq3/MgO/Co/Ru (thicknesses 20/12/3/30/10 nm, respectively). The active device area is lithographically defined by a window of approx. 500 x 500 nm² through an insulating aluminum oxide (AlOx) layer on top of the LSMO. All other active layers are consecutively deposited through different large-area shadow masks by thermal evaporation (Alq3), sputtering (MgO, Co, Ru) and e-beam evaporation (Ti/Au contacts). During the last step the organic is shielded by the shadow mask to avoid any radiation damage.[8]
Characterization is done in a $^4$He bath cryostat equipped with a 3D vector magnet. All results presented here were obtained at a temperature of 4.3 K. The resistance is measured at 10 mV bias (applied to the cobalt top contact) using a current amplifier. For RS effects short voltage pulses (200-500 ms) of up to ±2.5 V are applied, followed by a resistance measurement. For selected resistance values an I/V curve is taken and an MR scan is performed.

A total of 69 devices with nanosized active area on 15 different samples, each sample with respective different fabrication details were fabricated and characterized. Similar to Ref. 7 we achieve a yield of approx. 55% of working devices, while the other devices show an immeasurably high (>100 GΩ) or very low (<5 kΩ) resistance. This variation of the electrical device properties among a series of samples has recently been discussed[7] and indicates that tunneling through pinholes is the dominating transport mechanism. The presence of TMR and the absence of GMR are also confirmed by Hanle-measurements which are routinely performed on all devices.[6]
For more than half of the working devices it is possible to change the device resistance reversibly by applying voltage pulses of different height and polarity and all of these devices exhibit MR. Though strong variations in total resistance and MR appear from device to device as mentioned above the qualitative behavior is comparable and is described below using two examples.

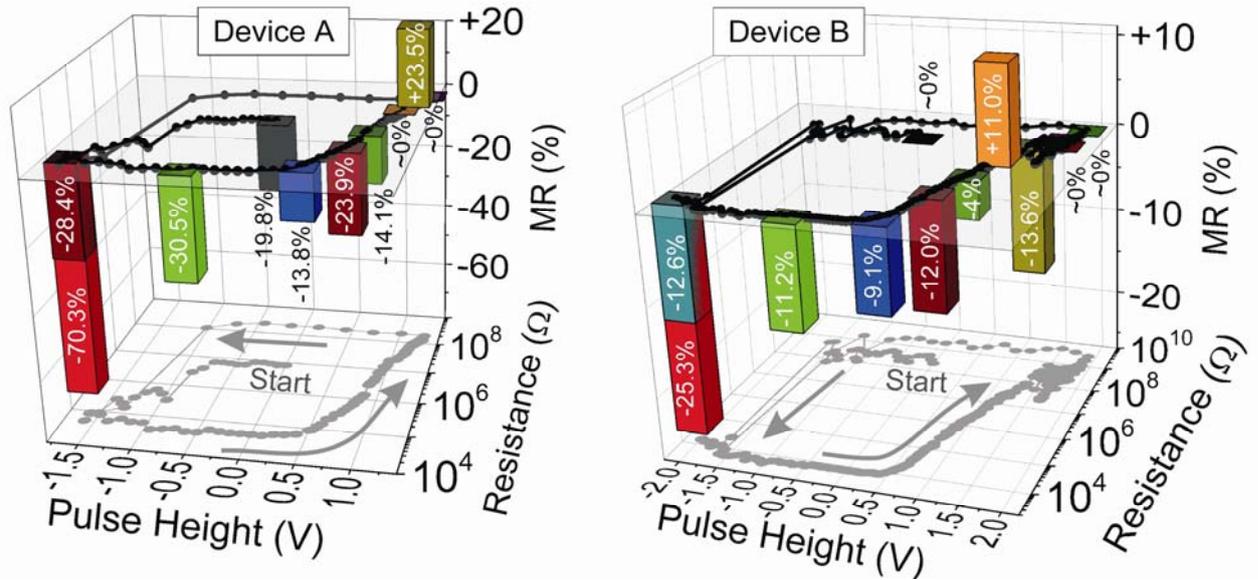

FIG. 2. Resistance versus pulse height (XY plane with its projection to the bottom plane) for two nanosized (500 x 500 nm²) perpendicular OSVs with $d_{Alq3} \approx 12$ nm. A closed loop over 3 decades of resistance (5 for device 'B') is possible. The bars indicate the maximum relative MR at selected points of the loop. The MR is always negative except for a single positive result close to the maximum resistance state (see Fig. 3 for more details). All measurements were done at 4.3 K.

Fig. 2 shows the changes of the device resistance and the relative MR for a representative device (Device A) when voltage pulses between -1.6 V and +1.3 V are applied. The initial resistance before any pulse (8.1 MΩ) only changes when $V_{pulse}$<-0.7 V. Further decrease of the pulse voltage down to -1.6 V results in a decrease of resistance by more than two decades (≈50 kΩ). From this point on the pulse voltage is increased. Massive increase of resistance of more than three decades can be observed for $V_{pulse}$>+0.5 V, resulting in 270 MΩ after $V_{pulse}$=+1.3 V. Decreasing the pulse voltage again shows no effect down to -1.0 V when the resistance decreases again in a sharp switching event, finally reaching the lowest state ($V_{pulse}$=-1.6 V). We thus obtain a closed hysteresis loop which can be repeatedly driven through nearly 4 decades of resistance. At the two threshold values of $V_{pulse}$ the resistance rises smoothly (≈+0.5 V) but decreases in an abrupt manner (≈-1.0 V). The loop was not extended further in order to avoid damage to the device by excessive voltages.

Another device 'B' is located on another sample, however, fabricated using the same parameters. Here the resistance change extends over more than 5 decades (right side of Fig. 2), although the general behavior is the same as for device 'A'. The corresponding threshold voltages are slightly different and the measured resistances exhibit a distinct fluctuation for higher $V_{pulse}$, nevertheless a closed loop can be measured.

At selected points of the hysteresis loop (marked with colored bars along the MR axis in Fig. 2) MR measurements have been carried out. The bars indicate the respective relative magnitude of the MR. Already here it is visible that the MR not only changes its amplitude but also its sign, much in contrast to results of Prezioso et al.[3,4] and Grünewald et al.[5] The corresponding full MR traces are plotted in Fig. 3. The value for the maximum MR for each trace relates to the height and sign of the corresponding bar with the same color in Fig. 2.

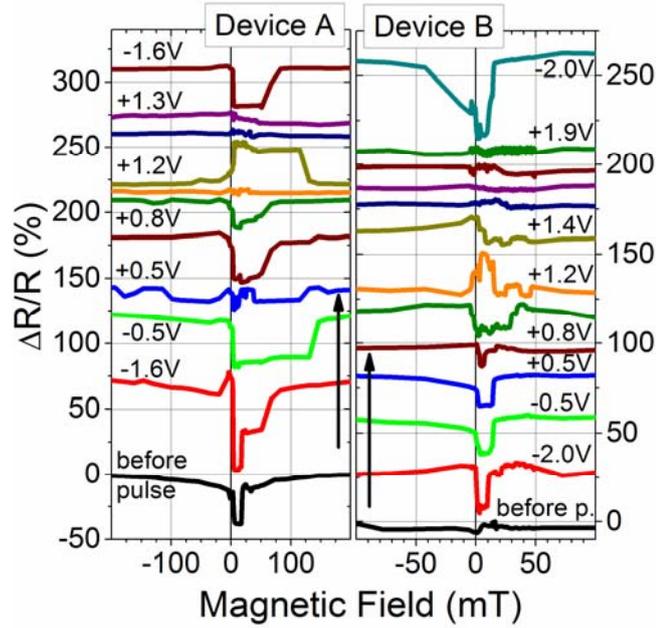

FIG. 3. MR traces of one magnetic field sweep direction for different voltage pulses (bottom to top) at T = 4.3 K for two devices 'A' and 'B', respectively. The MR traces are captured directly after resistance measurements which are depicted in Fig. 2 (according color code). The size, coercive fields and even the sign of the MR changes for both devices.

The black curve for device 'A' (bottom left in Fig. 3) is taken before any pulse is applied. The MR trace is typical for an OSV as it displays a negative MR (-19.8%) with two distinct resistance values and two clear coercive fields (+2 mT and +20 mT). The back sweep looks similar with reversed magnetic fields and is not shown here. Asymmetries in MR traces (compare to Fig. 1) often occur. They can also be seen elsewhere[1] and usually no clear explanation is given. In our devices they may for example be attributed to different pinning of domain walls between opposite magnetic field sweep directions, which could be related to the inhomogeneous Co top electrode or to a very sensitive probing of the complex density of states in the LSMO which depends on the exact direction of the magnetization with respect to the lattice and which may not vary uniformly during the reorientation process.

While the device resistance is changed by the voltage pulses (between -1.6 V and +1.3 V, bottom to top in Fig. 3 for device 'A') also all three features of the MR, namely the sign, the relative magnitude, and the coercive fields change. It should, however, be noted that all MR traces have a negative sign, except after $V_{pulse}$=+1.2 V where a pronounced positive MR signal of +23.5% can be observed. It is also noteworthy that in all MR traces the lower coercive field remains unchanged (≈2 mT) while the upper coercive field varies (17-140 mT). In some cases the switching at higher fields even occurs in several steps.

Starting with an MR of -20% (before any voltage pulse) it reaches -70% after the first pulse (-1.6 V). Successively a reduction (from -30% to <-4%) is observed until at +1.2 V even the sign is reversed (+23.5%) before the MR is finally reduced to almost zero. After closing the hysteresis loop and setting the device back to the initial low resistance state the MR also goes back to -28%, however, it then lacks the step-wise switching which could be observed in the beginning of the loop.

Qualitatively the same features as explained above can be observed for device 'B' (right side of Fig. 3) with two noticeable differences. Firstly, a MR trace can only be observed after an initial -2.0 V pulse and secondly, both coercive fields vary for the different states. It does not surprise that repeating the complete measurement cycles does not yield the exact same values of the MR, coercive fields and threshold voltages for the sign reversal as the preparation of each resistance state depends on the samples' history. Especially shape and magnitude of the MR can strongly depend on minor changes of the procedure while the overall qualitative behavior during similar cycles is identical. Indeed this directly supports our line of arguments about the presence of tunneling and its high sensitivity to any change of material properties.

We expect the RS in our device to originate from a tunnel barrier which is modified by a voltage pulse similar to the one described in Ref. 5. In order to analyze the barrier properties first I/V characterization is performed at those points of the hysteresis loop where also the MR is characterized. Small excitation voltages (±10 mV) are used in order to avoid modifications of the barrier during the measurement. The I/V curves are non-linear and the dI/dV curves exhibit a strongly parabolic character in the voltage range of approx. ±4 mV. This parabolic behavior is a typical indicator for tunneling processes as will be discussed below. The obtained curves, however, exhibit a certain asymmetry indicating that in contrast to Ref. 5 they are not caused and cannot be modeled by a simple rectangular barrier.

In the following explanation this is taken into account by considering two adjacent barriers with different origin and voltage dependence.

First we consider the tunneling through the Alq3 layer at pinholes which we do model by a fixed barrier for a given device and which is the major cause of TMR. But secondly our devices also exhibit RS which modulates the MR. This effect is well investigated by Grünewald *et al.*[5] and has successfully been modeled by a variable rectangular tunneling barrier created by oxygen vacancies at the Alq3/LSMO interface. These positive oxygen vacancies are formed at the LSMO/Alq3 interface. They are mobile and can easily be moved by electric fields. During a hysteresis loop vacancies and interstitials of oxygen ions are separated and redistributed within a certain depth from the interface which results in a variable tunnel barrier and causes the change in resistance which can be considered as a local metal-to-insulator transition.[9] It should be noted that the bulk of the LSMO electrode itself behaves fully metallic. Its in-plane resistance is in the range of a few kΩ and is slightly reduced during cool-down from room temperature to 4.3 K.

This second effect is implemented here by an additional second variable tunneling barrier in the same way as Grünewald *et al.* did for their devices.

Together with the aforementioned fixed Alq3 barrier we thus obtain a transport model of two adjacent tunnel barriers that we can simulate with a suitable fitting model, which is explained later.

While in Ref. 5 the variable barrier resulted in additional TAMR only, here we observe a much stronger effect. The features of the TMR now depend on the spin dependent coupling between the LSMO and the Co through the barrier. As the LSMO surface is part of the barrier its modification can change this coupling which can result in massive changes of the TMR, even in a sign reversal. Both barrier thickness[10] and electrode surface[11] are known to influence TMR with organic spacers in magnitude and sign. A modification below the LSMO surface can even lead to a modified magnetization reversal which may account for the observed changes in coercive field.

In order to implement the above explained transport model into an I/V fitting model with two adjacent tunnel barriers we use an expanded Simmons model[12] (with error corrected formulas). Each rectangular barrier is defined by the thickness, height, permittivity and charge carrier mass. For the first barrier all parameters are kept constant at reasonable values to account for the fixed Alq3 barrier for the different resistance states in one hysteresis loop (0.9 nm, 21.9 meV, 50 x $\varepsilon_0$ and 3.5 times the electron mass, respectively). For the second barrier at the surface of the LSMO, however, voltage pulses modify the barrier properties and cause the resistive switching. Therefore these parameters are kept free within adequate limits. As there is no global optimum for all nine fitting parameters any results need to be considered with caution and merely to extract trends rather than exact quantities.

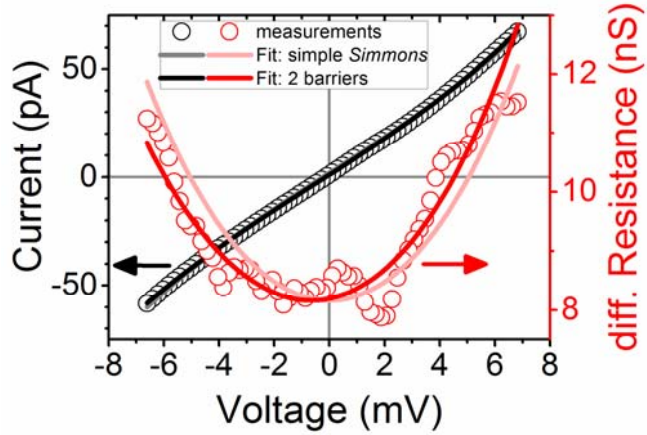

FIG. 4. Experimental data (circles) for I/V curve (black) and differential resistance (red) together with simulated curves for a simple Simmons fit (light colored curves) and for the two-barrier-model (dark colored curves) after a +1.2 V pulse for device 'A'. The latter one respects for asymmetries and achieves a better compliance with the measurement.

In Fig. 4 the experimental I/V and differential resistance curves are displayed after a +1.2 V pulse together with a simple Simmons fit (light colored) and a fit for the two-barrier-model (dark color). Obviously a better compliance is possible for the two-barrier-model especially as the asymmetry of the I/V characteristics is taken into account.

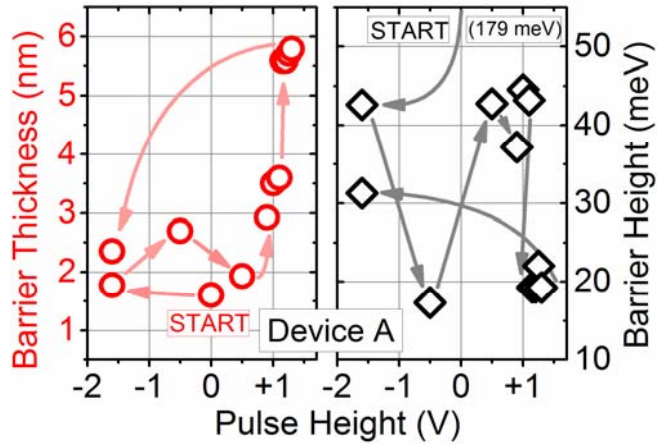

FIG. 5. Thickness and height of the second tunnel barrier versus the applied pulse voltage for device 'A'. Values have been extracted from data fitting using the expanded (and corrected) Simmons model[12] with two barriers. Parameters of the first barrier are fixed, most notably the thickness at 0.89 nm. The variable thickness and height of the second barrier (1.5 - 5.8 nm) are strongly influenced by different voltage pulses. Arrows indicate the sequence of the measurement in the resistance hysteresis loop beginning at 'start'.

For the variable barrier the most significant parameter which influences the transmission is the barrier thickness, for which we plotted the calculated value of each fit for different voltage pulses in Fig. 5. In our model allowed values are between 0.1 - 20 nm, while barrier height, effective mass of charge carriers and dielectric constant are allowed to vary within the ranges of 9.4 meV - 4 eV, 10 - 100 x $\varepsilon_0$ and 3 - 4 times the electron mass, respectively. According to the fit the barrier thickness changes from 1.5 nm to 5.8 nm and shows a clear trend of increasing thickness with higher pulse voltages. For the resetting pulse (-1.6 V) the barrier thickness also resets to a lower value of about 2.3 nm. These results support the theory about a variable tunnel barrier in the LSMO in addition to the fixed tunnel barrier of the Alq3 pinholes.

Although the explanation given above is plausible other explanations shall briefly be discussed. The Co contact itself is a possible candidate, however, nothing indicates a contribution from this side while we know that the LSMO contact can exhibit resistive switching.[13–16] In addition RS in oxides has often been described[17–21] while for metallic cobalt RS has not been reported. Modification of conducting filaments in the Alq3 can be excluded because we already know that the MR is based on tunneling. Other modifications of the Alq3 cannot be completely excluded, though no suitable mechanism comes to mind. In reference to the explanation of the MR and RS effects provided in Refs. 3 and 4 our alternative explanation is entirely evident as the basic magnetoresistive functionality of our devices differs considerably.

In addition our theory can explain the finding that an MR effect in device 'B' could only be measured after applying a negative voltage pulse by the need to first sufficiently create and separate oxygen vacancies and interstitials to realize a suitable barrier for the occurrence of tunneling assisted MR effects as described in Ref. 5 again supporting the presence of a similar mechanism.

Furthermore, recent measurements on similar devices containing metal free phthalocyanine ($H_2Pc$) as spacer layer which will be published elsewhere show very similar results in terms of MR and RS properties supporting that the effect originates from the electrode rather than from properties of the organic semiconductor.

In summary we have fabricated nanosized perpendicular organic spin valve devices which exhibit a unique interplay of resistive switching and magnetoresistance effects up to now primarily observed in multiferroic tunnel junctions. Transport occurs locally through pinholes providing a tunneling path through the organic material. The MR thus originates only from TMR and possibly from small contributions from TAMR. The tunnel barrier consists of two parts, namely the organic semiconductor in the pinhole (fixed barrier) and a tunnel barrier induced by oxygen vacancies at the LSMO surface which can be modified by a voltage pulse leading to resistive switching and modifications of the MR (variable barrier).

These devices demonstrate the potential multifunctionality of LSMO in transport structures and present the opportunity for multi-state memory because in contrast to Prezioso *et al.*[3,4] the MR persists in different resistive switching states. A major challenge, however, for applications will be to harness the MR which is currently based on pinholes and thus difficult to control.

**Acknowledgments**
This work was supported by the European Commission within the 7FP project HINTS (Project No. NMP-CT-2006-033370) and by the DFG in the SFB 762.